\documentclass[reprint,superscriptaddress,amssymb,aps, pra, twocolumn,graphicx]{revtex4-1}

\usepackage{graphicx}
\usepackage{dcolumn}
\usepackage{bm}
\usepackage{hyperref}

\usepackage{amsmath} 
\usepackage{diagbox}
\usepackage{mathrsfs}
\usepackage[cmintegrals]{newtxmath}
\usepackage[mathlines]{lineno}

\begin{document}


\title{Proximity-encirclement of exceptional points in a multimode optomechanical system}

\author{Zheng Fan}
\affiliation{Department of Physics, State Key Laboratory of Low-Dimensional Quantum Physics, Tsinghua University, Beijing 100084, China}

\author{Dan Long}
\email{longd20@mails.tsinghua.edu.cn}
\affiliation{Department of Physics, State Key Laboratory of Low-Dimensional Quantum Physics, Tsinghua University, Beijing 100084, China}

\author{Xuan Mao}
\affiliation{Department of Physics, State Key Laboratory of Low-Dimensional Quantum Physics, Tsinghua University, Beijing 100084, China}

\author{Guo-Qing Qin}
\affiliation{Department of Physics, State Key Laboratory of Low-Dimensional Quantum Physics, Tsinghua University, Beijing 100084, China}

\author{Min Wang}
\email{wangmin@baqis.ac.cn}
\affiliation{Beijing Academy of Quantum Information Sciences, Beijing 100193, China}

\author{Gui-Qin Li}
\affiliation{Department of Physics, State Key Laboratory of Low-Dimensional Quantum Physics, Tsinghua University, Beijing 100084, China}
\affiliation{Frontier Science Center for Quantum Information, Beijing 100084, China}

\author{Gui-Lu Long}
\affiliation{Department of Physics, State Key Laboratory of Low-Dimensional Quantum Physics, Tsinghua University, Beijing 100084, China}
\affiliation{Beijing Academy of Quantum Information Sciences, Beijing 100193, China}
\affiliation{Frontier Science Center for Quantum Information, Beijing 100084, China}

\date{\today}

\begin{abstract}

 Dynamic encircling a second-order exception point (EP) exhibit chiral state transfer, while there
 is few research on dynamic encircling multiple and higher-order EPs. Here, we study proximity-encirclement of the EPs in a multimode optomechanical system to understand the closed path evolution of high-order non-Hermitian systems. The optomechanical system has three types of situations about EPs: the system has no EP, a pair of second-order EPs, and a third-order EP. The dynamical behavior of the system's dependence on the initial state, orientation, and velocity of the loop, the variance in the starting point of the loop, as well as the number and order of EPs encircled by the loop have been investigated in the process of state transfer. The results show that chiral or non-reciprocal state transfer can be realized when the loop encircling a second-order EP with different radius. Only chiral state transfer occurs when encircling two second-order EPs. Moreover, chiral and non-reciprocal state transfer can happen in a single loop encircling a third-order EP. The phenomena about encircling the EPs in a multimode optomechanical system provides another means for manipulating state transfer in higher-order non-Hermitian systems.

\end{abstract}


\maketitle


\section{INTRODUCTION \label{introduction}}

In recent years, exceptional points (EPs), where the eigenvalues and eigenvectors coalesce simultaneously in non-Hermitian system\cite{RN1.1.1,RN1.1.2,RN1.2.1,RN1.2.2}. It has many applications attached much attention\cite{RN1.2.2,RN1.3.1,RN1.3.2,RN1.3.3,RN1.3.4,RN1.3.5,RN1.3.6,RN1.3.7,RN1.3.8,RN1.3.9,RN1.3.10,RN1.3.11} such as sensing\cite{RN1.3.1,RN1.3.2,RN1.3.3,RN1.3.4,RN1.3.5,RN1.3.6}, state conversion\cite{RN1.3.7,RN1.3.8} and phonon laser\cite{RN1.3.9,RN1.3.10,RN1.3.11}. Due to the flexibility of system control, the optomechanical systems\cite{RN1.4.1,RN1.5.1,RN1.5.2,RN1.5.3,RN1.5.4,RN1.5.5,RN1.5.6,RN1.5.7,RN1.5.8,RN1.5.9,RN1.5.10,RN1.5.11,RN1.5.12,RN1.5.13} have become a promising and convenient platform for manipulating EPs to approach state transfer\cite{RN1.3.8}. The optomechanical system has become an ideal research platform for fundamental physics to study electromagnetically induced transparency\cite{RN1.6.1,RN1.6.2}, optical information storage\cite{RN1.7.1}, high efficiency frequency conversion\cite{RN1.8.1}, and phonon cooling\cite{RN1.9.1}. Given that the Hamiltonian has topological structure around the EP\cite{RN1.10.1}, state conversion occurs when the system is manipulated to encircle the EP in the parameter space\cite{RN1.11.1}. It is found that state transfer can be chiral\cite{RN1.3.7,RN1.12.1} or non-reciprocal\cite{RN1.3.8}, which has been realized in multiple systems such as circuits\cite{RN1.13.1}, waveguides\cite{RN1.12.1}, and plasmonics\cite{RN1.14.1,RN1.14.2}.

The majority of previous literatures have focused on EP of second-order systems\cite{RN1.3.7,RN1.12.1}, where chiral state transfer happens when a second-order EP has been encircled\cite{RN1.10.1}. This is important for understanding the closed-path evolution of non-Hermitian systems\cite{RN2.1.1}, which will help with system control\cite{RN2.2.1} and EP-based photonics device design\cite{RN2.3.1}. However, a multimode systems may have multiple and higher-order EPs, results a unique phenomena of encircling of the EPs\cite{RN2.4.1,RN2.5.1,RN2.5.2}.

In this study, we investigate the encircling of two second-order EPs and a third-order EP in a multimode optomechanical system, which is composed of two optical modes and a mechanical mode. The results show that in this system, in addition to the initial state, orientation, and velocity of the loop, the variance in the starting point of the loop, the number and order of EPs encircled by the loop, will also have an impact on the evolutionary results. When the loop encircles only one second-order EP, the radius of the loop determines whether there is chiral or non-reciprocal state transfer. For the case where the loop encircles two second-order EPs, there are different chiral state transfer results with different starting points. Moreover, when a third-order EP is encircled, chiral and non-reciprocal state transfer can be realized depending the starting points.These results will help the system control as well as the design of higher-order EP photonic devices. Because, non-reciprocal and chiral evolution results can be achieved on a single trajectory, this will be helpful for energy transfer and state manipulation.

The remainder of this article is organized as follows: In Sec.\ref{MODEL AND HAMILTONIAN}, we describe the basic model, the Hamiltonian and the solutions of the system. We discuss the dynamic evolution of the system in Sec.\ref{DYNAMICALLY}. Discussion and  conclusions are presented in Sec.\ref{Conclusion}. In Appendix, we  presents the distribution of the exceptional points of the system, the construction of the new projection vectors, the instantaneous amplitudes of eigenstates during the loop evolution that encloses two second-order EPs and a third-order EP.

\section{MODEL AND HAMILTONIAN \label{MODEL AND HAMILTONIAN}}

\begin{figure}
	\centering
	\includegraphics[width=\linewidth]{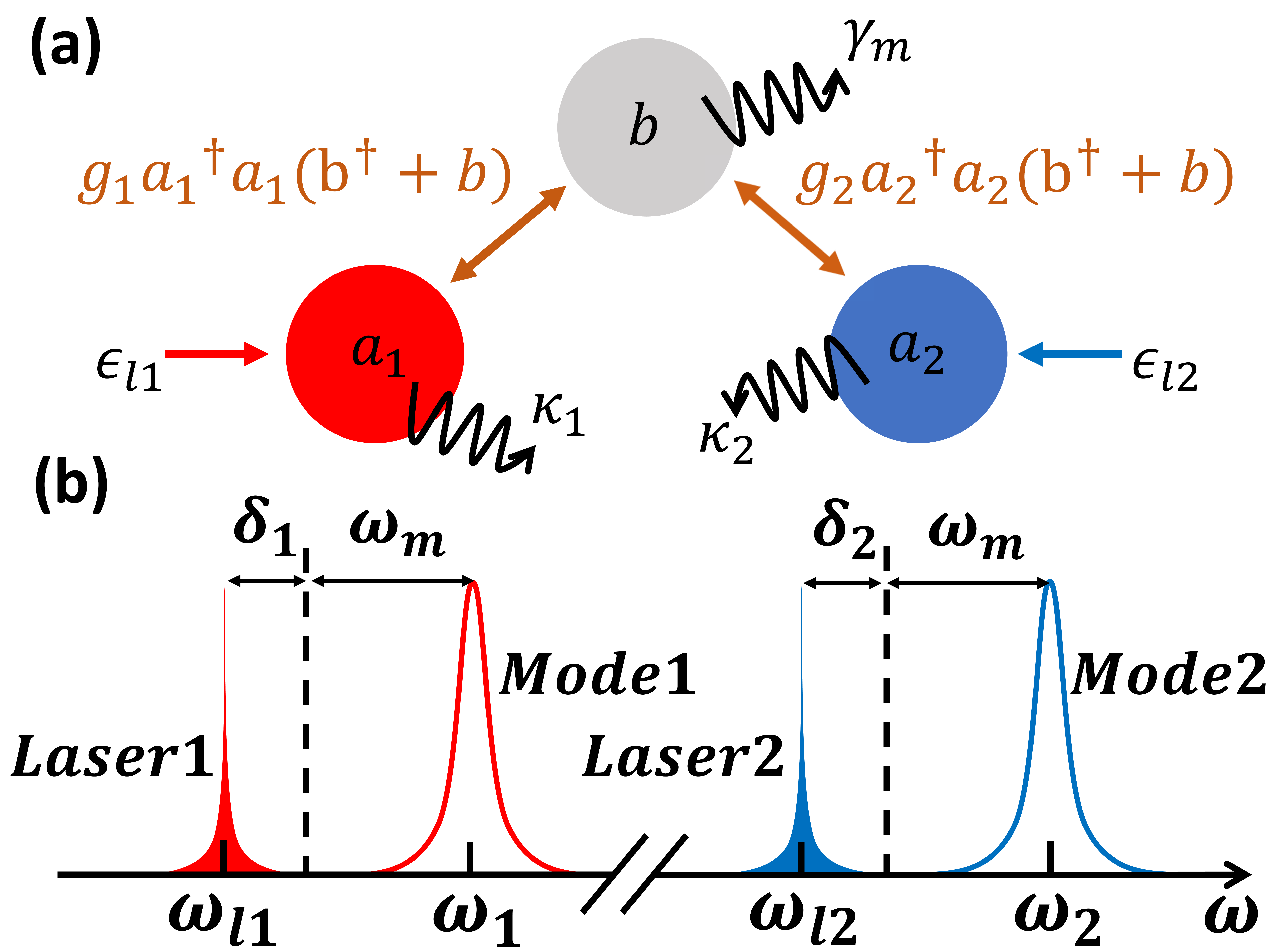}
	\caption{(a) Schematic of the optomechanical system composed of two optical modes ($a_1$ and $a_2$) and one mechanical mode ($b$). The two optical modes are coupled with the mechanical modes simultaneously. (b) Frequency spectrum of the two optical systems and two strong power lasers with red detunings.}
	\label{model}
\end{figure}

A schematic representation of the proposed system is shown in Fig. \ref{model}(a), which is composed of two optical modes and one mechanical mode. The two optical modes were driven by strong lasers with central frequencies of $\omega_{l1}$ and $\omega_{l2}$, which is exhibited in Fig. \ref{model}(b). The damping (gaining) rates of the optical modes were $\kappa_j$, and the mechanical mode was $\gamma$. The mechanical mode was simultaneously and dispersively coupled with the two optical modes. The system Hamiltonian can be described by

\begin{align}
	H &= H_{{\rm free}} + H_{{\rm int}} + H_{{\rm drive}},
	\label{equation 1}
\end{align}
where 

\begin{figure*}
	\centering
	\includegraphics[width=\linewidth]{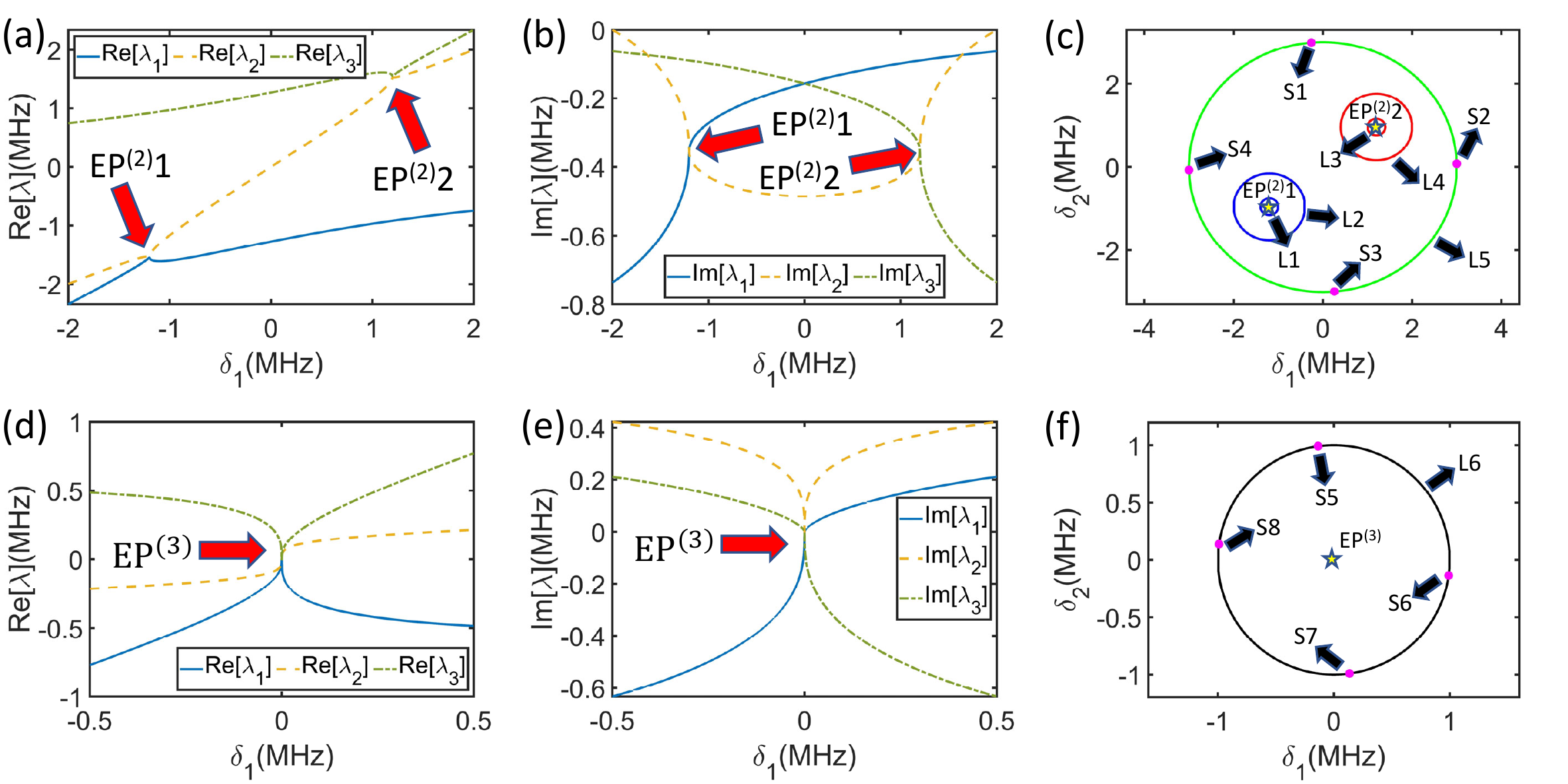}
	\caption{(a) Real part Re$[\lambda]$ and (b) imaginary part Im$[\lambda]$ of the effective Hamiltonian eigenvalues as functions of the detuning of the laser $\delta_1$, and instantaneously $\delta_2$ varying linearly with $\delta_1$. (c) Parametric space of the system with second-order EPs. There are five different loops marked as L1-L5 in this subgraph. For loop L5, there are four different starting points marked as S1-S4 in this subgraph. (d) Real part Re$[\lambda]$ and (e) imaginary part Im$[\lambda]$ of the effective Hamiltonian eigenvalues as functions of the detuning $\delta_1$ with $\delta_2$ fixed at 0. (f) Parametric space of the system with a third-order EP. The loop marked as L6 with four different starting points S5-S8 in this subgraph. The values of the  determined parameters and the location of the EPs were given in the Table.\ref{tb1}}
	\label{hamiltonian}
\end{figure*}

\begin{align}
	H_{{\rm free}} =& \omega_1{a_1}^\dagger a_1+\omega_2{a_2}^\dagger a_2+\omega_mb^\dagger b, \nonumber \\
	H_{{\rm int}} =& g_1{a_1}^\dagger a_1(b^\dagger+b)+g_2{a_2}^\dagger a_2(b^\dagger+b), \nonumber\\
	H_{{\rm drive}} =&i\sqrt{\kappa_{ex1}}\epsilon_{l1}e^{-i\omega_{l1}t}{a_1}^\dagger+i\sqrt{\kappa_{ex2}}\epsilon_{l2}e^{-i\omega_{l2}t}{a_2}^\dagger+{\rm H.c.},\nonumber\\
	\label{equation 2}
\end{align}
here $\omega_{j}$ denotes the resonance frequency of the $j^{th}$ optical mode, and $\omega_m$ describes the resonance frequency of the mechanical mode. $a_j$ and $b$ are the annihilations of the $j^{th}$ optical mode and mechanical mode, respectively; $g_j$ and $\kappa_{\rm{exj}}$ describe the single-photon optomechanical coupling rate and coupling damping rate, respectively; and $\epsilon_{lj}$ is the power of the $j^{th}$ laser.

In the interaction picture of the driving field, the Hamiltonian is rewritten as 

\begin{align}
	H &= \omega_mb^\dagger b-\sum_{j=1,2}[\Delta_j{a_j}^\dagger a_j-g_j{a_j}^\dagger a_j(b^\dagger+b)] \nonumber \\ 
	&+i\sqrt{\kappa_{ex1}}\epsilon_{l1}{a_1}^\dagger+i\sqrt{\kappa_{ex2}}\epsilon_{l2}{a_2}^\dagger,
	\label{equation 3}
\end{align}
where $\Delta_j=\omega_{lj}-\omega_j$ denotes the detuning between the optical modes and corresponding driving fields. It is convenient to solve the nonlinear Heisenberg equations if the Hamiltonian above is linearized. The annihilation of the optical modes can be changed by $a_j\rightarrow\overline{a_j}+a_j$. The linearized Hamiltonian is converted to:

\begin{align}
	H=&\omega_mb^\dagger b-\sum_{j=1,2}[\Delta_j{a_j}^\dagger a_j-G_j({a_j}^\dagger+a_j)(b^\dagger+b)],
	\label{equation 4}
\end{align}
where $G_j=g_j\overline{a_j}$ describes the effective optomechanical coupling strength between the mechanical mode and cavity mode $j$, and $\overline{a_j}=\sqrt{\kappa_{exj}}\epsilon_{lj}/(-i\Delta_j+\frac{\kappa_j}{2})$ is the average annihilation. 

We consider the case in which both optical modes are driven under red sidebands. For convenience, the detuning can be rewritten as  $\delta_j=-\Delta_j-\omega_m$. Under the condition $\omega_m\gg(\delta_j,G_j)$ and the rotating-wave approximation, the Hamiltonian can be written as 

\begin{equation}
	H_A=\sum_{j=1,2}[\delta_j{a_j}^\dagger a_j+G_j({a_j}^\dagger b+a_jb^\dagger)],
\end{equation}
 in the interaction picture of $\omega_m(b^\dagger b+\sum_{j=1,2}{a_j}^\dagger a_j)$. In this case, the Langevin equation of the system is 
 
 \begin{align}
 	i\dot{\vec {v}}_A(t)=&M_A\vec{v}_A(t)+i\sqrt{K}\vec{v}_{in}^{A}(t),\nonumber \\ \vec{v}_A(t)=&\left[a_1(t),b(t),a_2(t)\right]^T
 \end{align}
 where $\vec{v}_{in}^{A}(t)=\left[a_{in}^{(1)}(t),b_{in}(t),a_{in}^{(2)}(t)\right]^T$, 

\begin{align}
	\mathcal{M_A}=&
	\begin{pmatrix}
		\delta_1-i\kappa_1/2 & G_1 & 0 \\
		G_1 & -i\gamma_m/2 & G_2\\
		0 & G_2 & \delta_2-i\kappa_2/2
	\end{pmatrix},
	\label{equation 5}
\end{align}
and $K=diag\left[\kappa_1,\gamma_m,\kappa_2\right]$. As a result of $\gamma_m\ll(\kappa_1,\kappa_2)$ for a general optomechanical system, the mechanical loss $\gamma_m$ can be ignored. Thus, the effective Hamiltonian of our system is reduced to,

\begin{align}
	\mathcal{H}=&
	\begin{pmatrix}
		\delta_1-i\kappa_1/2 & G_1 & 0 \\
		G_1 & 0 & G_2\\
		0 & G_2 & \delta_2-i\kappa_2/2
	\end{pmatrix},
	\label{equation 6}
\end{align}

\begin{figure*}
	\centering
	\includegraphics[width=\linewidth]{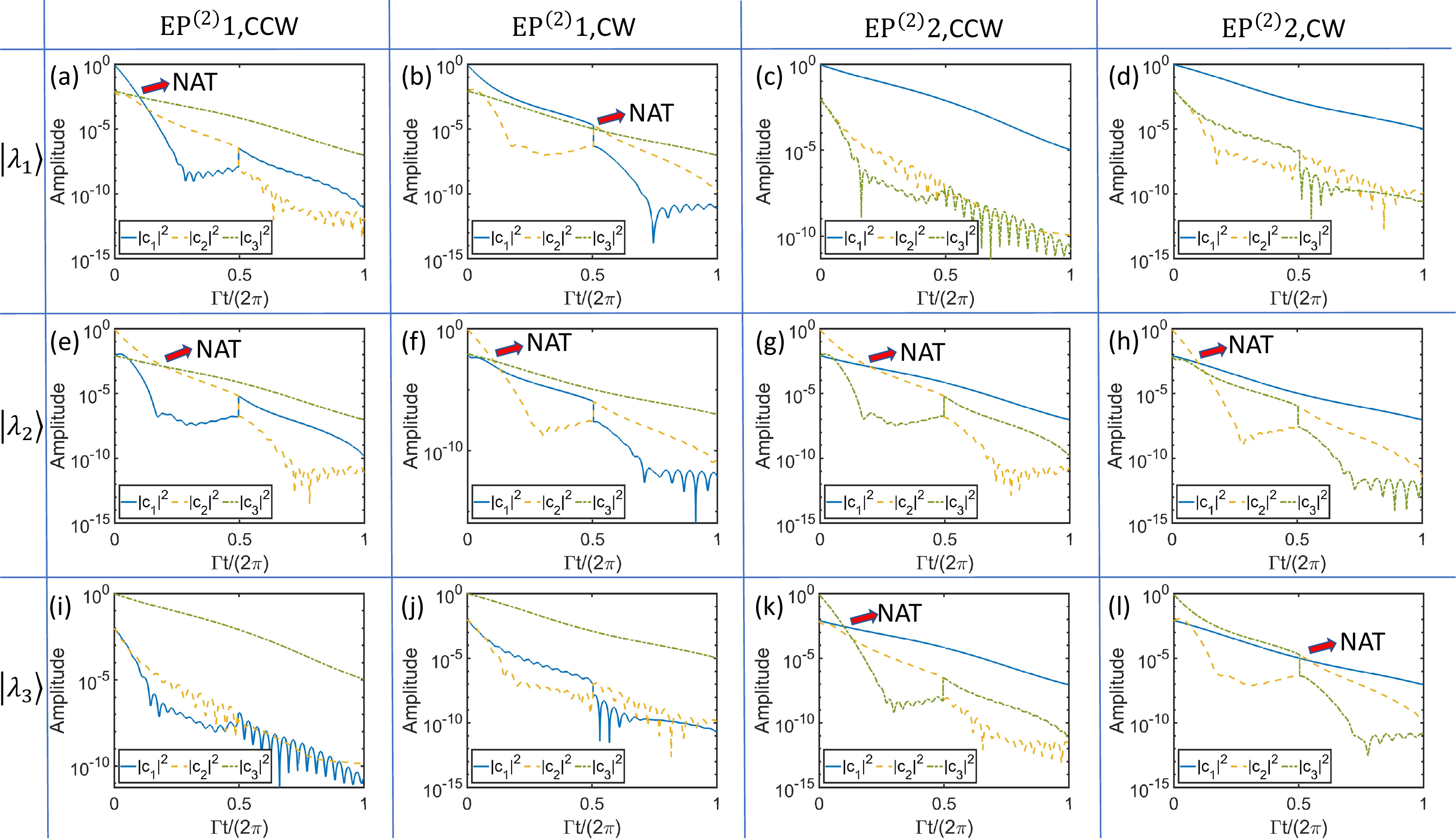}
	\caption{Instantaneous amplitudes of eigenstates during the loop evolution that encloses a second-order EP, where the radius of the loop is 0.2G. The blue solid line describes the amplitude of $\lambda_1$, the yellow dashed line describes the amplitude of $\lambda_2$ and green dashed dot line describes the amplitude of $\lambda_3$. The rows of the subgraph indicate the distinct initial states, the columns of the subgraph represent the different orientations of the circle or the different second-order EP encircled. }
	\label{encircling1}
\end{figure*}

\setlength{\tabcolsep}{0.3mm}{\begin{table}[]
		\caption{The values of determined parameters and the location of the EPs about the Fig.\ref{hamiltonian}(c) and (f). In the table below, G=1MHz}
		\begin{tabular}{|c|c|c|}
			\hline
			The cases                                                                        & \begin{tabular}[c]{@{}c@{}}The values of the\\ determined parameters\end{tabular} & \begin{tabular}[c]{@{}c@{}}The location\\ of the EPs\end{tabular}                                        \\ \hline
			\begin{tabular}[c]{@{}c@{}}The system \\ has two\\ second-order EPs\end{tabular} & \begin{tabular}[c]{@{}c@{}}$G_1=G_2=1{\rm G}$,\\ $\kappa_1=1{\rm G}$,$\kappa_2=-0.2{\rm G}$\end{tabular}            & \begin{tabular}[c]{@{}c@{}}${\rm EP}^{(2)}1$:\\$\delta_1=-1.1995{\rm G}$,\\   $\delta_2=-0.9576{\rm G}$\\ \\ ${\rm EP}^{(2)}2$:\\$\delta_1=1.1995{\rm G}$,\\   $\delta_2=0.9576{\rm G}$\end{tabular} \\ \hline
			\begin{tabular}[c]{@{}c@{}}The system\\ has a\\ third-order EP\end{tabular}      & \begin{tabular}[c]{@{}c@{}}$G_1=G_2=\sqrt{2}{\rm G}/2$,\\ $\kappa_1=1{\rm G}$,$\kappa_2=-1{\rm G}$\end{tabular}            & ${\rm EP}^{(3)}$:$\delta_1=\delta_2=0$                                                                                            \\ \hline
		\end{tabular}
		\label{tb1}
\end{table}}

Though both the eigenvalues and the eigenvectors for a three-dimensional non-Hermitian system may be obtained analytically, however the analytic eigenvalues would be exceedingly complicated and are not easy to analyze the distribution of EPs. Instead, we use numerical solutions for the investigation. There are three possible outcomes for this system regarding EPs:the system has no EP, a pair of second-order EPs, and a third-order EP. Whether the system has second-order EPs can be identified using Cardano’s formula method's discriminant\cite{RN3.1.1}. The distribution of exceptional points of the system is exhibited in Appendix.\ref{AppendixA}. A third-order EP can only exist in this system when it possesses PT symmetry.

Fig.\ref{hamiltonian}(a)-(b) depict the real and imaginary parts of the effective Hamiltonian with two second-order EPs, whereas (d)-(e) depict the real and imaginary parts of the effective Hamiltonian with a third-order EP. As can be seen in Fig.\ref{hamiltonian}(a)-(b), $\lambda_1 =\lambda_2$ for ${\rm EP}^{(2)}1$ are the same, and $\lambda_2 =\lambda_3$ for ${\rm EP}^{(2)}2$. The three eigenvalues are identical at the third-order EP, as shown in Fig.\ref{hamiltonian}(d)-(e). The locations of the second-order EPs are shown in Fig.\ref{hamiltonian}(c) and the third-order EP in Fig.\ref{hamiltonian}(f), with the second-order EPs being centrosymmetric around the third-order EP in parametric space of the detunings $\delta_1$ and $\delta_2$.

\section{DYNAMICALLY ENCIRCLING THE EXCEPTIONAL POINTS \label{DYNAMICALLY}}

\begin{figure*}
	\centering
	\includegraphics[width=\linewidth]{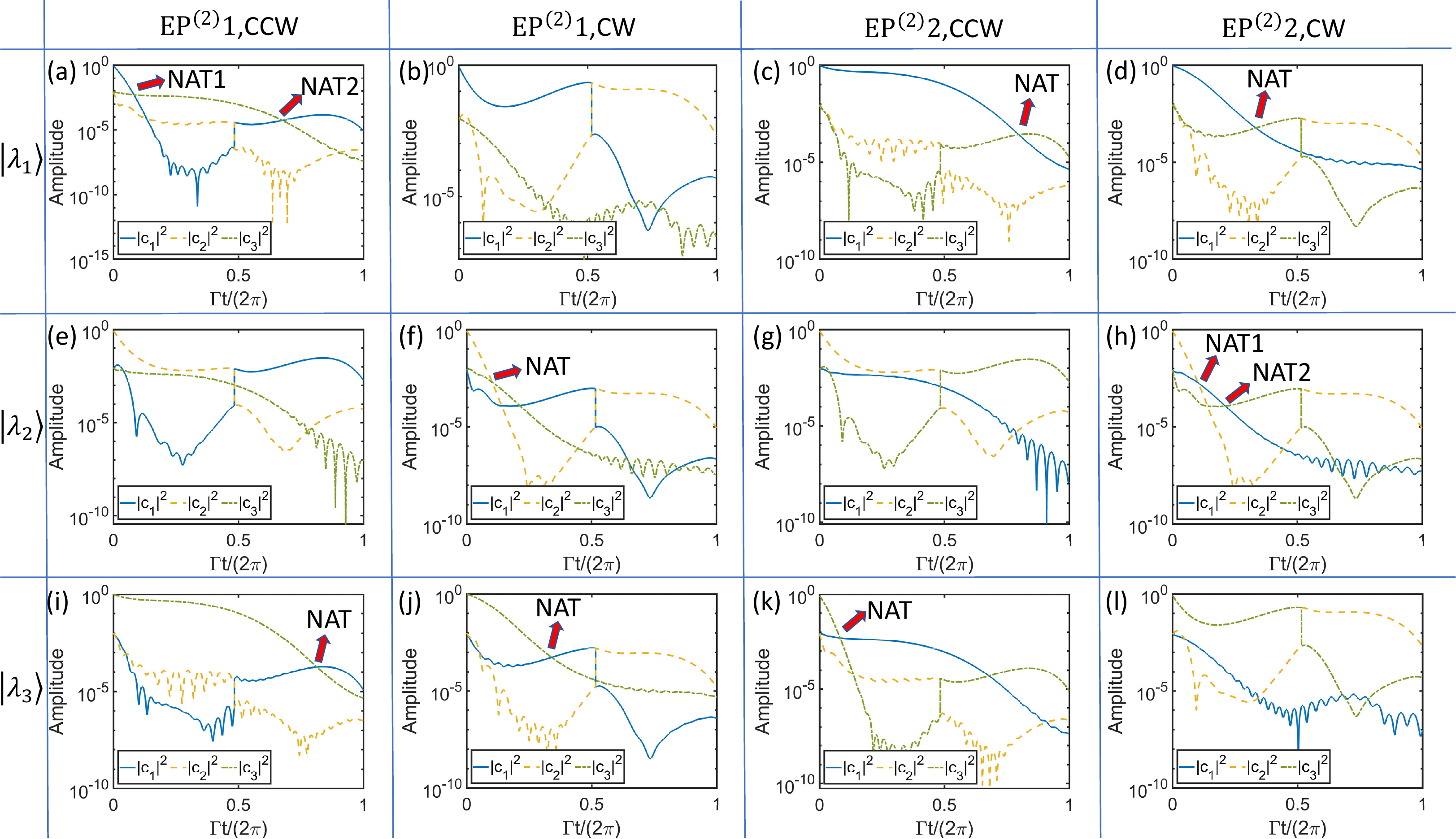}
	\caption{Instantaneous amplitudes of eigenstates during the loop evolution that encloses a second-order EP, where the radius of the loop is 0.8G. The blue solid line describes the amplitude of $\lambda_1$, the yellow dashed line describes the amplitude of $\lambda_2$ and green dashed dot line describes the amplitude of $\lambda_3$. The rows of the subgraph indicate the distinct initial states, the columns of the subgraph represent the different orientations of the circle or the different second-order EP encircled.}
	\label{encircling2}
\end{figure*}

\begin{figure}
	\centering
	\includegraphics[width=\linewidth]{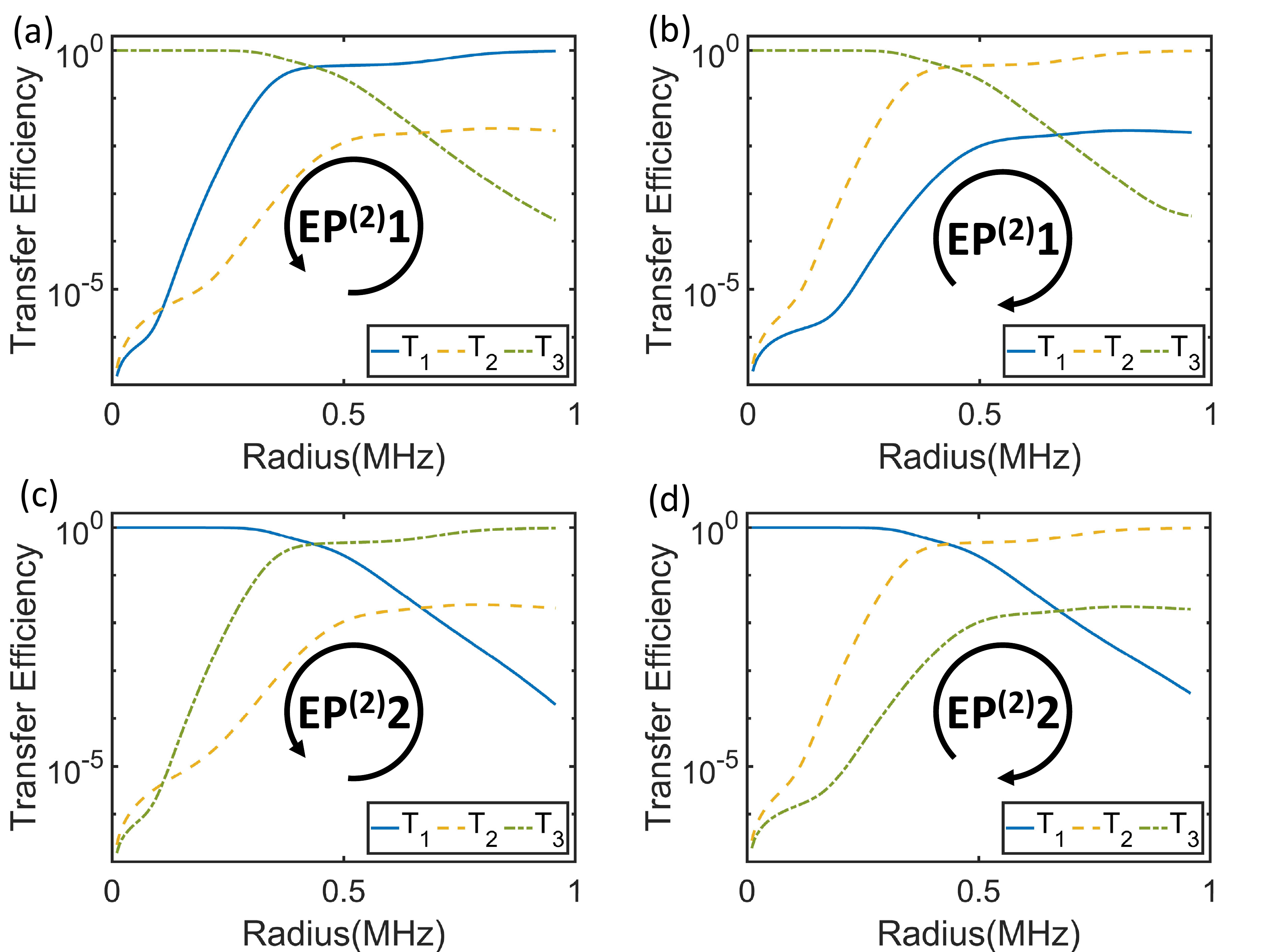}
	\caption{The energy transfer efficiency T as a function of the loop radius R when the loop encircles a second-order EP. Encircling ${\rm EP}^{(2)}1$ corresponds to (a) and (b), while encircling ${\rm EP}^{(2)}2$ corresponds to (c) and (d). The encircling direction of (a) and (c) is CCW and the direction of (b) and (d) is CW. }
	\label{transfer1}
\end{figure}

\begin{figure*}
	\centering
	\includegraphics[width=\linewidth]{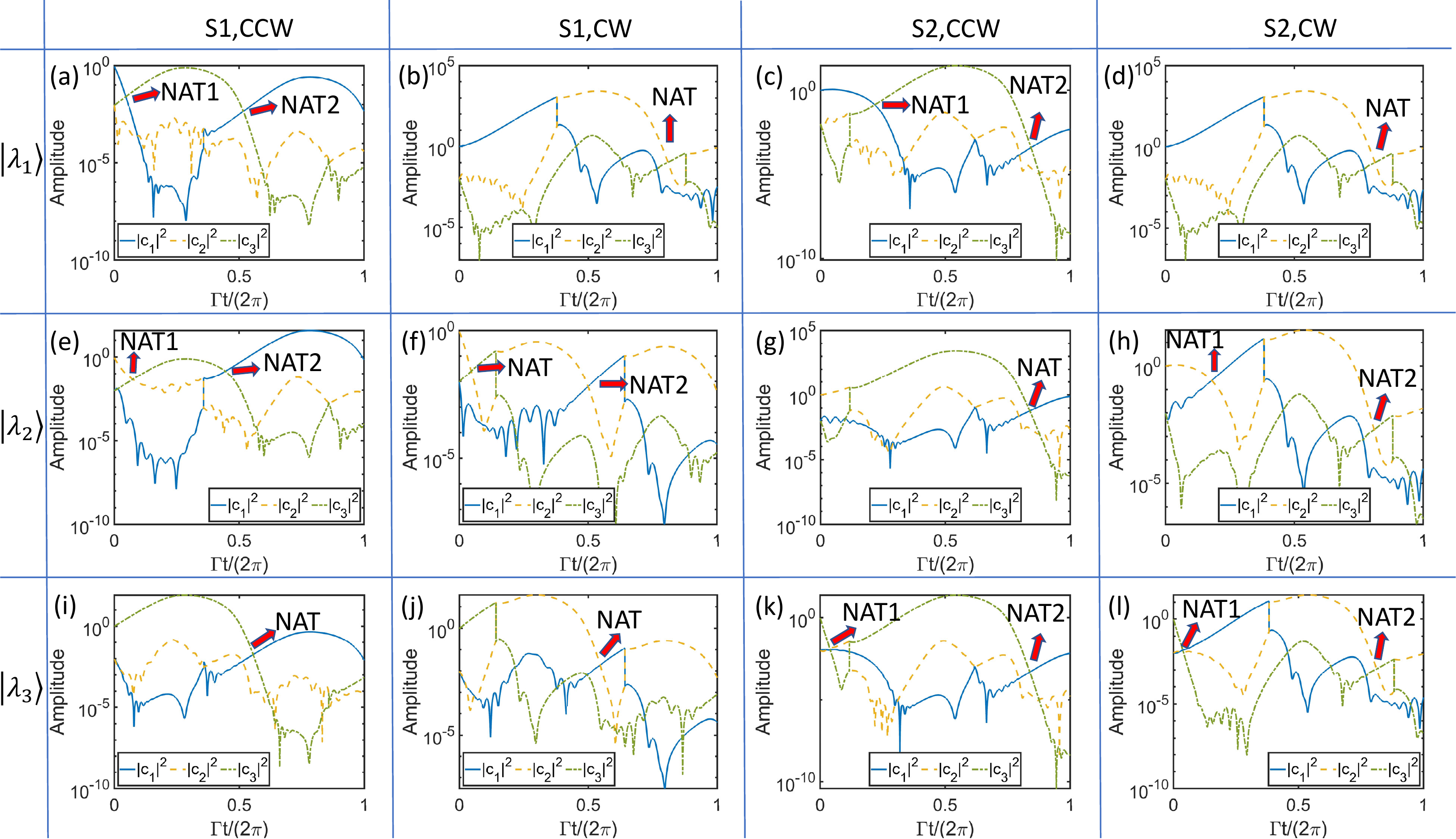}
	\caption{Instantaneous amplitudes of eigenstates during the loop evolution that encloses two second-order EPs, where the radius of the loop is 3G. The blue solid line describes the amplitude of $\lambda_1$, the yellow solid line describes the amplitude of $\lambda_2$ and green solid line describes the amplitude of $\lambda_3$.The rows of the subgraph indicate the distinct initial states, the columns of the subgraph represent the different orientations of the circle or the different starting point of the circle.}
	\label{encircling3}
\end{figure*}

The Heisenberg equation $dA/dt=i/h[H,A]$ controls the dynamical evolution of the optomechanial system. It is possible to develop a controlled dynamical evolution around EPs in parameter space by concurrently manipulating $\delta_1$ and $\delta_2$. The time-dependent state of the system can be represented as 

\begin{equation}
	|\psi(t)\rangle=c_1(t)|\lambda_1(t)\rangle+c_2(t)|\lambda_2(t)\rangle+c_3(t)|\lambda_3(t)\rangle,
\end{equation}
 where $c_1$, $c_2$, and $c_3$ signify the instantaneous amplitudes of these three eigenvalues, if $|\lambda_1(t)\rangle$, $|\lambda_2(t)\rangle$, and $|\lambda_3(t)\rangle$ are designated as the instantaneous eigenvectors of the system. The instantaneous eigenvalue amplitudes cannot be directly derived by projecting the state $|\psi(t)\rangle$ onto the eigenstates $|\lambda_i(t)\rangle$ because the system is non-Hermitian. In order to acquire the amplitudes, new vectors can be established by
\begin{equation}
	\begin{pmatrix}
		|l_1(t)\rangle\\
		|l_2(t)\rangle\\
		|l_3(t)\rangle
	\end{pmatrix}
	=
	\begin{pmatrix}
		1 & \alpha_1 & \beta_1\\
		\beta_2 & 1 & \alpha_2\\
		\alpha_3 & \beta_3 & 1
	\end{pmatrix}
	\cdot
	\begin{pmatrix}
		|\lambda_1(t)\rangle\\
		|\lambda_2(t)\rangle\\
		|\lambda_3(t)\rangle
	\end{pmatrix}
	\label{equation7}
\end{equation}
and the amplitudes can be derived by projecting the state $|\psi(t)\rangle$ of the system to these new vectors $|l_i(t)\rangle$. The specific calculations and values for $\alpha$ and $\beta$ are shown in Appendix.\ref{AppendixB}.

It is straightforward to learn about evolution by concentrating at the instantaneous amplitudes of the eigenvalues. The simplest function of the loop is expressed as 

\begin{align}
	\delta_1(t)=&\alpha_1 \delta_1^{EP}+\beta_1 \delta_1^{EP}{\rm cos}(\Gamma t+\phi_0),\\
	\delta_2(t)=&\alpha_2 \delta_2^{EP}+\beta_2 \delta_2^{EP}{\rm sin}(\Gamma t+\phi_0),
\end{align}
where $\Gamma$ describes the velocity of evolution and its sign describes the orientation of the loop. $\Gamma>0$ corresponds to a counterclockwise (CCW) loop, whereas $\Gamma<0$ corresponds to a clockwise (CW) loop. $\alpha_i$ determines the location of the loop center, and $\beta_i$ determines the radius of the loop. The phase of the starting point is determined by $\phi_0$ when we determine that the time interval is $[0, 2\pi/|\Gamma|]$. The starting point is always selected where the two imaginary parts of the eigenvalues coalesce. Without loss of generality, it is assumed that most of the energy is initially concentrated on one of the eigenstates.

The dynamical evolution of third-order systems is more complex than that of second-order systems. As compared to the system with only the second-order EP, the present system contains two additional situations one with two second-order EPs and with a third-order EP. The dynamical evolution of a third-order system also depends on the radius of the loop, the order, and the number of the EPs it encircles, in addition to the initial state, orientation and location of the loop, which will depend the result of the dynamical evolution in the second-order systems.

\subsection{DYNAMICAL ENCIRCLING A SECOND-ORDER EXCEPTIONAL POINT}

Fig.\ref{encircling1} shows the instantaneous amplitudes of the eigenvectors when encircling a second-order EP with a small radius loop L1 and L3, which is shown at Fig.\ref{hamiltonian}(c). $\alpha_1=\pm1.1995{\rm G}$, $\alpha_2=\pm0.9576{\rm G}$ and $\beta_1=\beta_2=0.2{\rm G}$ ensure the loops enclose the EP with a small radius. $\Gamma=0.1{\rm G}$ ensures that the evolution is adiabatic. It is obvious that there is a state-flip when $\Gamma t/(\pi)=1$ in all subgraphes, owing to the topological structure of the imaginary parts of the eigenvalues around the second-order EP. Because of the time-dependent Hamiltonian, the state $|\psi(t)\rangle$ will not precisely be the eigenstates $|\lambda_i(t)\rangle$ of the instantaneous Hamiltonian during the evolution. If the state first propagates on the higher-loss sheet, the evolution will be unstable, and if the evolution is adiabatic, the state will gradually propagate to the lower-loss sheet. When two of the amplitudes coincide like $c_1(t_1)=c_2(t_1)$ as in Ref.\cite{RN2.6.1}, it is the confirmation of the occurrence of non-Hermiticity-induced nonadiabatic transition (NAT)\cite{RN2.6.1,RN2.6.2} at time $t=t_1$. When the loop encloses the ${\rm EP}^{(2)}1$, the state always transfer into the $|\lambda_3\rangle$ whatever the initial state and the loop orientation are, which is shown in Fig.\ref{encircling1}(a)-(b), (e)-(f) and (i)-(j). No matter what the initial state and loop orientation are, when the loop encloses the ${\rm EP}^{(2)}2$, the state always transfers into the $|\lambda_1\rangle$, as demonstrated in Fig.\ref {encircling1}(c)-(d), (g)-(h), and (k)-(l) respectively. It is clear that when the radius is small, the dynamic evolution leads to non-reciprocity and that the consequences of the evolution are unaffected by the orientation of the loop. 

However, the dynamical evolution results show chirality when the loop L2 and L4 encircle a second-order EP with a large radius, which is shown in Fig.\ref{encircling2}. $\alpha_1=\pm1.1995{\rm G}$, $\alpha_2=\pm0.9576{\rm G}$ and $\beta_1=\beta_2=0.8{\rm G}$ ensure the loops enclose the EP with a large radius. $\Gamma=0.1{\rm G}$ ensures that the evolution is adiabatic. When the loop is oriented CCW around the ${\rm EP}^{(2)}1$, the state always changes to $|\lambda_1\rangle$; when the loop is oriented CW, the state always evolves to $|\lambda_2\rangle$. The state always progresses to $|\lambda_3\rangle$ when the loop is oriented CCW around the ${\rm EP}^{(2)}2$, but the state always changes to $|\lambda_2\rangle$ when the loop is oriented CW. It is clear that when the radius is large enough, the dynamic evolution leads to chirality , which is dependent of the loop orientation. 

When the loop radius is small enough, the evolution show non-reciprocal, whereas the evolution is chiral when the loop radius is large enough. Changing the loop radius leads to different imaginary parts of the Hamiltonian where the loop goes through, and different imaginary parts of the Hamiltonian cause different losses of the eigenstates, resulting in the diverse outcomes as shown in Fig.\ref{encircling1} and Fig.\ref{encircling2}. To quantify the effect of loop radius on evolutionary outcomes, we examine the energy transfer during the loop with different radius. The energy transfer efficiencies are defined by
\begin{figure*}
	\centering
	\includegraphics[width=\linewidth]{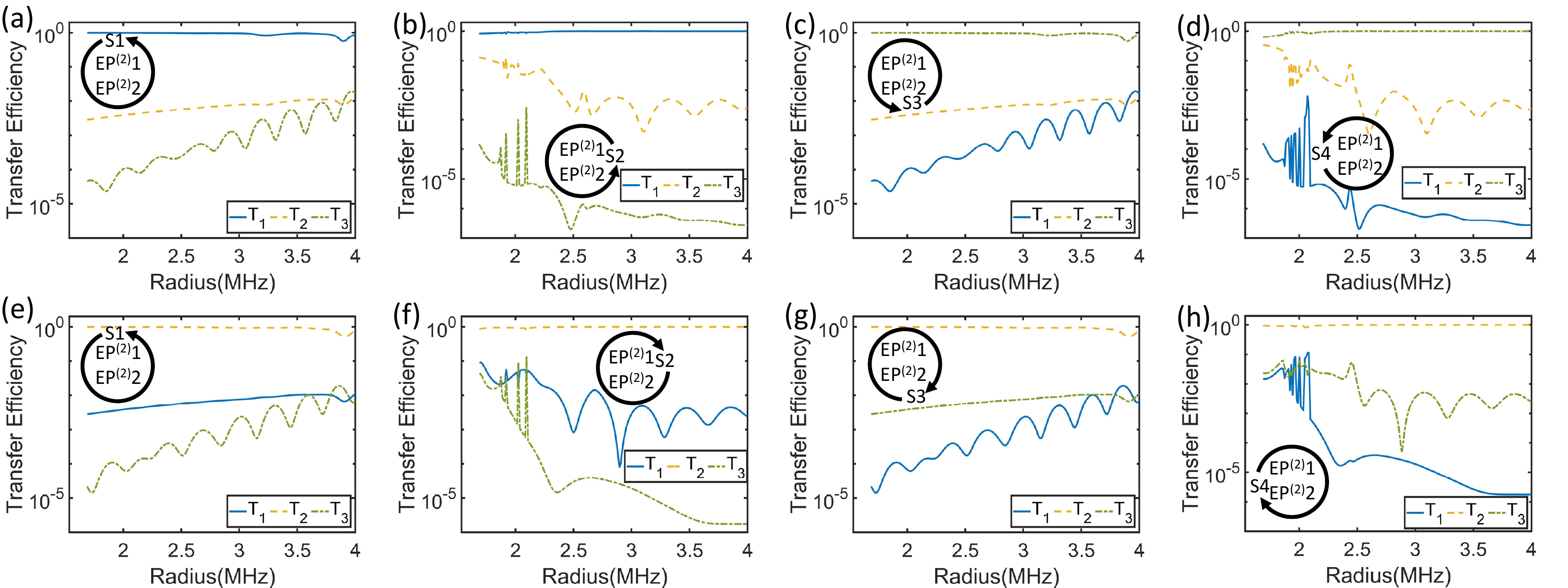}
	\caption{The energy transfer efficiency T as a function of the loop radius R when the loop encircles two second-order EPs. Starting point S1-S4 corresponds to (a)-(d) with orientation is CCW, respectively. Starting point S1-S4 corresponds to (e)-(h) with orientation is CW, respectively.}
	\label{transfer2}
\end{figure*}

\begin{equation}
	\begin{pmatrix}
		T_1\\
		T_2\\
		T_3
	\end{pmatrix}
	=
	\begin{pmatrix}
		0 & \frac{|c_{12}|^2}{2} & \frac{|c_{13}|^2}{2}\\
		\frac{|c_{21}|^2}{2} & 0 & \frac{|c_{23}|^2}{2}\\
		\frac{|c_{31}|^2}{2} & \frac{|c_{32}|^2}{2} & 0
	\end{pmatrix}
	\cdot
	\begin{pmatrix}
		\frac{1}{|c_{11}|^2+|c_{21}|^2+|c_{31}|^2}\\
		\frac{1}{|c_{12}|^2+|c_{22}|^2+|c_{32}|^2}\\
		\frac{1}{|c_{13}|^2+|c_{23}|^2+|c_{33}|^2}
	\end{pmatrix}
	\label{equation8},
\end{equation}
where $c_{ij}$ define the amplitudes of the eigenvalues $|\lambda_i(t)\rangle$ after the loop evolution and most of the energy contribute to the eigenvalue $|\lambda_j(t)\rangle$ before the evolution. By measuring the energy transfer efficiency, Fig.\ref{transfer1} confirms the findings of Fig.\ref{encircling1} and Fig.\ref{encircling2}. It is obvious that the lines in Fig.\ref{transfer1}(a) and (c) have the same shape, only the color is distinct, which is the same in Fig.\ref{transfer1}(b) and (d). This is because ${\rm EP}^{(2)}1$ and ${\rm EP}^{(2)}2$ are symmetric about the origin in the parameter space and the structure of the two second-order EPs is the same. In summary, the evolution results are non-reciprocal for small radius while chiral for large radius.

\subsection{DYNAMICAL ENCIRCLING TWO SECOND-ORDER EXCEPTIONAL POINTS}

There are four starting points for the loop when the loop consists of two second-order EPs. These starting points are the places where the imaginary parts of $|\lambda_1(t)\rangle$ and $|\lambda_2(t)\rangle$ coincide, or the places where the imaginary parts of $|\lambda_2(t)\rangle$ and $|\lambda_3(t)\rangle$ coincide. However, when two of the imaginary parts of the eigenvalues coincide, the other imaginary part of the eigenvalue is the largest or the smallest. So there are four starting points named as S1-S4, which are shown in Fig. \ref{hamiltonian}(c), and the correspond cases are shown in Table.\ref{t2}. $\alpha_1=\alpha_2=0$ and $\beta_1=\beta_2=3{\rm G}$ ensure the loops enclose two second-order EPs. $\Gamma=0.1{\rm G}$ ensures that the evolution is adiabatic. 

From Fig.\ref{encircling3} and Fig.\ref{appendix2} in Appendix.\ref{AppendixC}, when the loop L5 encloses two second-order EPs, the evolution results are dependent of the starting points and show chirality. When the starting point is S1 or S2 and the orientation is CCW, the state always transfers into $|\lambda_1\rangle$ whereas the state always changes into $|\lambda_2\rangle$ when the orientation is CW. When the starting point is S3 or S4 and the orientation is CCW, the state always transfers into $|\lambda_3\rangle$ whereas the state always changes into $|\lambda_2\rangle$ when the orientation is CW. When alternative starting points are chosen, the relative size of the imaginary parts of the Hamiltonian of the starting points influence the evolutionary outcomes. As a result, the chiral evolution produces separate results for the S1-S2 and S3-S4 starting points, which correspond to different eigenstates after the evolution.

As can be seen from Fig.\ref{transfer2}, when the path contains two second-order EPs, the size of the radius R does not change the relative magnitude of the energy transfer efficiency, i.e., it does not change the chiral evolution result. Moreover, when the orientation is CW, the state always transfer into $|\lambda_2\rangle$ whatever the starting points, whereas the orientation is CCW, the state transfer into $|\lambda_1\rangle$ and $|\lambda_3\rangle$ for the starting points are S1-S2 and S3-S4, respectively.

\subsection{DYNAMICAL ENCIRCLING A THIRD-ORDER EXCEPTIONAL POINT}

There are four optional starting points for the loop when the loop L6 encloses the third-order EP, as shown in Fig. \ref{hamiltonian}(f). The definition of the starting points in loop L6 are the same as those of loop L5. These four starting points are named S5-S8 as shown in Fig. \ref{hamiltonian}(f) and the correspond cases are shown in Table.\ref{t2}. $\alpha_1=\alpha_2=0$ and $\beta_1=\beta_2=1{\rm G}$ ensure the loop encloses the third-order EP.  $\Gamma=1{\rm G}$ ensures that the evolution is adiabatic. The instantaneous amplitudes of eigenstates are exhibited in Appendix.\ref{AppendixD}.

\setlength{\tabcolsep}{0.5mm}{\begin{table}[]
		\caption{The case corresponding to the starting points S1-S8 of loops L5-L6.}
		\begin{tabular}{|c|c|c|}
			\hline
			\begin{tabular}[c]{@{}c@{}}The starting\\ point\end{tabular} & \begin{tabular}[c]{@{}c@{}}Which two\\ imaginary parts of\\ eigenvalues coincide\end{tabular} & \begin{tabular}[c]{@{}c@{}}The other\\ imaginary part of\\ the eigenvalue:\\ Largest/Smallest\end{tabular} \\ \hline
			S1 and S5                                                          & $|\lambda_1\rangle$ and $|\lambda_2\rangle$                                                                                              & Largest                                                                                                    \\ \hline
			S2 and S6                                                          & $|\lambda_1\rangle$ and $|\lambda_2\rangle$                                                                                              & Smallest                                                                                                   \\ \hline
			S3 and S7                                                           & $|\lambda_2\rangle$ and $|\lambda_3\rangle$                                                                                              & Largest                                                                                                    \\ \hline
			S4 and S8                                                           & $|\lambda_2\rangle$ and $|\lambda_3\rangle$                                                                                              & Smallest                                                                                                   \\ \hline
		\end{tabular}
		\label{t2}
\end{table}}

\begin{figure*}
	\centering
	\includegraphics[width=\linewidth]{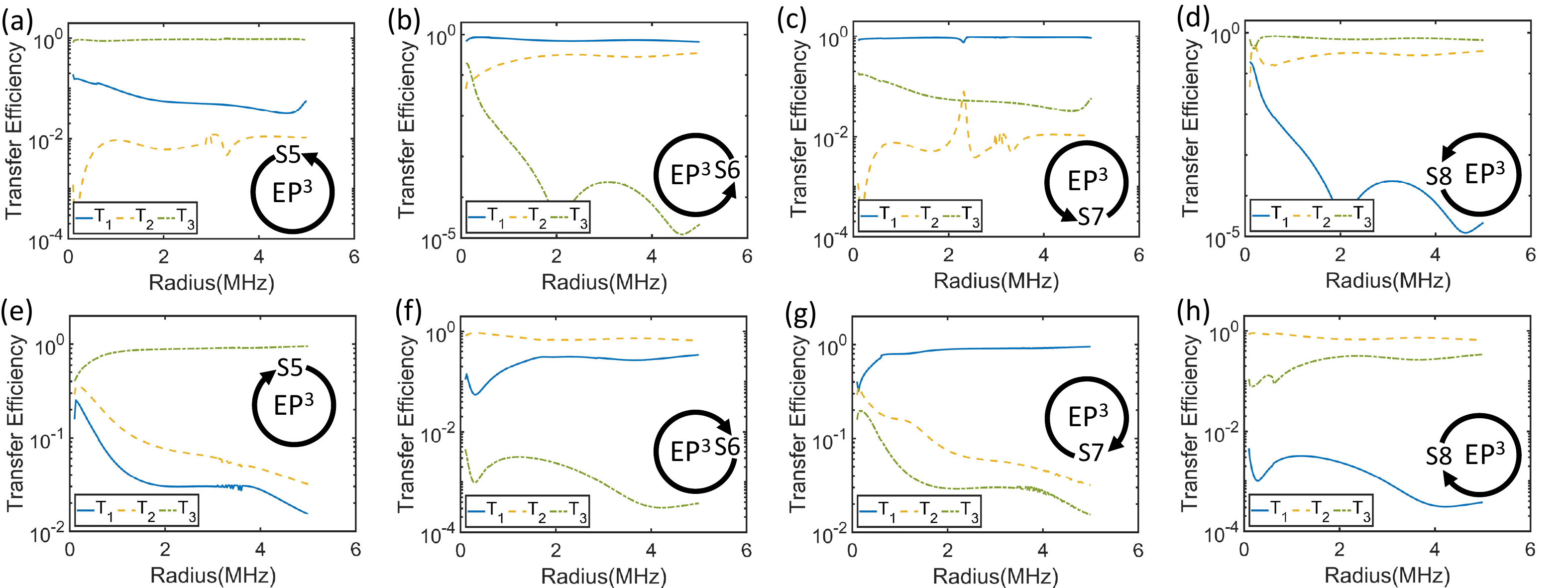}
	\caption{The energy transfer efficiency T as a function of the loop radius R when the loop encircles the third-order EP. Starting point S5-S8 corresponds to (a)-(d) with orientation is CCW, respectively. Starting point S1-S4 corresponds to (e)-(h) with orientation is CW, respectively.}
	\label{transfer3}
\end{figure*}

Irrespective of radius of the loop, the evolution outcomes are non-reciprocal when the starting points are S5 and S7, as shown in Fig.\ref{transfer3}. The $|\lambda_3\rangle$ always results from evolution when the beginning point is S5, but the $|\lambda_1\rangle$ always results from evolution when the starting point is S7. Regardless of the loop radius, the evolution outcome is chiral when the beginning points are S6 and S8, respectively. When the starting point is S6, the evolution leads to $|\lambda_1\rangle$ and $|\lambda_2\rangle$ depending on the direction is CCW or CW, while the starting point is S8, the evolution leads to $|\lambda_3\rangle$ and $|\lambda_2\rangle$ depending on the direction is CCW or CW.

On the loop around the third-order EP, there are four starting points where two of the imaginary parts of the Hamiltonian coincide, and the relative magnitudes of the three eigenvalue imaginary parts at these points are inconsistent. The evolutionary outcomes are impacted by the respective sizes of the three eigenvalue imaginary parts when these sites are chosen as starting points. Because of this, the evolution result of surrounding the third-order EP are strongly influenced by the choice of starting points.

\section{Conclusion \label{Conclusion}}

To summarize, we used an optomechanical system with two optical modes and one mechanical mode to study the encircling evolution around EPs such a system may not have an EP, two second-order EPs, and a third-order EP depending on system parameters including optically driven coupling and losses of the optical modes. A chiral energy transformation takes place around a second-order EP when it is surrounded by a large radius as opposed to a non-reciprocal energy transformation by a small radius. Depending on the starting points, different chiral energy transformations take place while encircling two second-order EPs. Two starting points when surrounding a third-order EP correspond to non-reciprocal evolution result, while the other two starting points correspond chiral evolution outcome. The evolution results will be related to the radius size only when encircling a second-order EP. Our findings enrich the understanding of the closed path evolution of high order non-Hermitian systems, which will be beneficial to system control and the design of high order EP-based photonics devices. While non-reciprocal and chiral energy conversion can be accomplished on only a loop of the system, our results will be more conducive to the creation of energy conversion across multimode, offering a new perspective for the study of system manipulation.

\begin{acknowledgments}
	
This work was supported by the National Natural Science Foundation of China (61727801 and 62131002), National Key Research and Development Program of China (2017YFA0303700), Special Project for Research and Development in Key Areas of Guangdong Province (2018B030325002), Beijing Advanced Innovation Center for Future Chip (ICFC), and Tsinghua University Initiative Scientific Research Program.
	
\end{acknowledgments}

\appendix

\section{Distribution of exceptional points of the system \label{AppendixA}}

In order to solve the eigenvalues and eigenstates of the third-order non-Hermitian system, it is necessary to solve the equivalent system of cubic equations with complex coefficients. For a cubic equation $ax^3+bx^2+cx+d=0$ with any coefficients, the discriminant 

\begin{align}
	\Delta=4ac^3-b^2c^2-18abcd+27a^2d^2+4b^3d
	\label{equation 9}
\end{align}
can be used to determine whether there exist two identical solutions. When this discriminant is zero, the system can be considered to have a pair of second-order EPs. In general, the discriminant for detuning is a quadratic equation, therefore calculating the equation with detuning $\delta_2$ as the independent variable for a discriminant of zero yields four complex solutions, the imaginary part of which may be ascribed to the loss of the second optical mode $\kappa_2$.

The loss of the system's first optical mode is restricted to $\kappa_1=1G$, while the coupling strength of the system's two optical and mechanical modes is limited to the same $G_1=G_2$. Traversing the detuning  $\delta_1$ produces all complicated solutions given $G_1$ and $\kappa_2$. The range of $\kappa_2$ for which the second-order EP system exists for a given $G_1$ may be found using these complicated solutions.

\begin{figure}
	\centering
	\includegraphics[width=\linewidth]{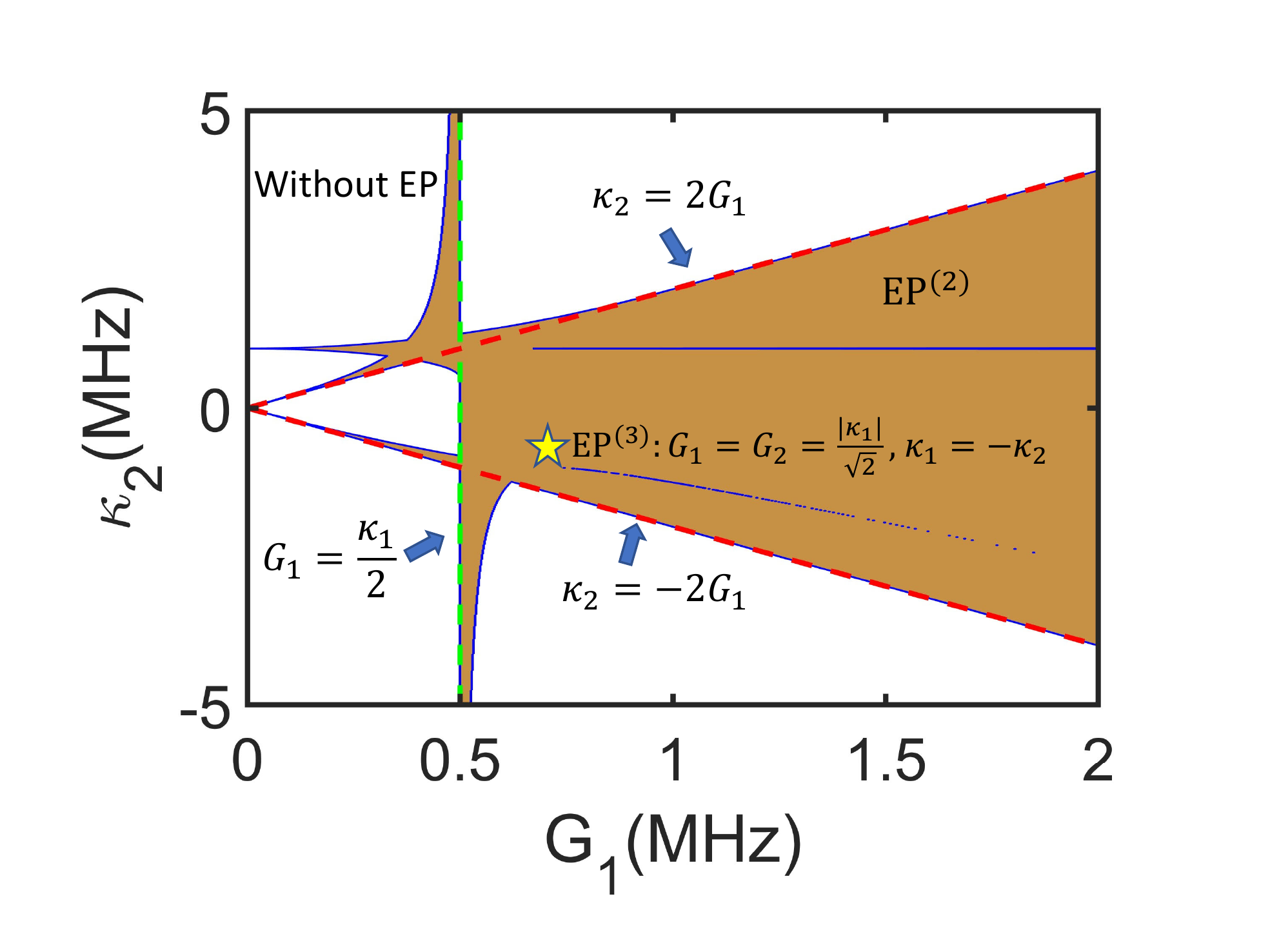}
	\caption{Existence of phase diagram for second-order EP.}
	\label{appendix1}
\end{figure}

Fig. \ref{appendix1} depicts the phase diagram for the presence or absence of second-order EP. The lack of EP is shown by white areas, whereas the presence of second-order EP is represented by copper-colored areas. The copper-colored region is separated into two large sections, $G_1<0.5$ and $G_1>0.5$. The first optical and mechanical modes couple weakly at $G_1<0.5$, i.e., $G_1<\kappa_1/2$, and the copper-colored area may be split into three tiny parts at this moment. These are the region extended by $\kappa_2\approx1$, and the two regions extended by $\kappa_2=\pm2G_1$ from $\kappa_2=0$, respectively. The top region is a second-order EP region created by the two optical modes, while the bottom two are second-order EP regions formed by the mechanical mode and the second optical mode.

Another large region, $G_1>0.5$, i.e., $G_1>\kappa_1/2$, where the first optical and mechanical modes couple strongly. At this case, the only method to determine the presence of second-order EP is to find for a strong coupling between the second optical mode and the mechanical mode. Therefore, in this large region, $\kappa_2=\pm2G_1$ determines whether or not there is a second-order EP. There is second-order EP when $|\kappa_2|<2G_1$, and there is no second-order EP when $|\kappa_2|>2G_1$.

\section{Construction of the new projection vectors \label{AppendixB}}

In Sec.\ref{DYNAMICALLY}  it is shown that the effective Hamiltonian is non-Hermitian so that the instantaneous amplitudes of the eigenstates cannot be obtained directly by projecting the state onto the eigenstates. By constructing the new projection vectors by means of Eq.\ref{equation7}, it is possible to obtain the instantaneous amplitudes. 

\begin{figure*}
	\centering
	\includegraphics[width=\linewidth]{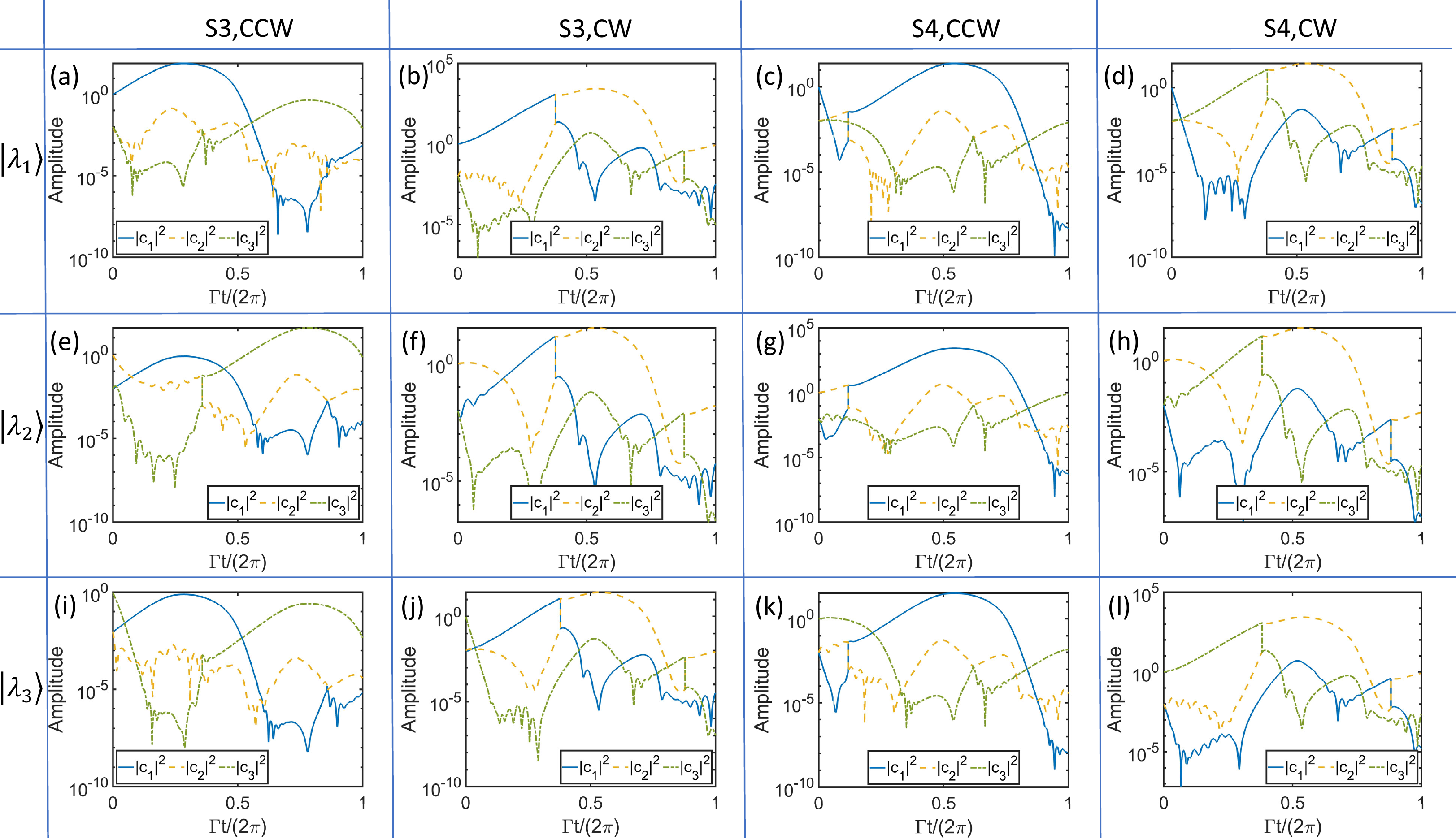}
	\caption{Instantaneous amplitudes of eigenstates during the loop evolution that encloses two second-order EPs, where the radius of the loop is 3G. The blue solid line describes the amplitude of $\lambda_1$, the yellow solid line describes the amplitude of $\lambda_2$ and green solid line describes the amplitude of $\lambda_3$.The rows of the subgraph indicate the distinct initial states, the columns of the subgraph represent the different orientations of the circle or the different starting point of the circle. }
	\label{appendix2}
\end{figure*}

The instantaneous state of the system is $|\psi(t)\rangle=c_1(t)|\lambda_1(t)\rangle+c_2(t)|\lambda_2(t)\rangle+c_3(t)|\lambda_3(t)\rangle$ and the first new projection vector is $|l_1(t)\rangle=|\lambda_1(t)\rangle+\alpha_1|\lambda_2(t)\rangle+\beta_1|\lambda_3(t)\rangle$. Projecting $|\psi(t)\rangle$ onto $|l_1(t)\rangle$ yields

\begin{align}
	\langle|l_1(t)|\psi(t)\rangle&= c_1+c_2\langle|l_1(t)|l_2(t)\rangle+c_3\langle|l_1(t)|l_3(t)\rangle\nonumber\\
	+&c_1{\alpha_1}^\ast\langle|l_2(t)|l_1(t)\rangle+c_2{\alpha_1}^\ast+c_3{\alpha_1}^\ast\langle|l_2(t)|l_3(t)\rangle\nonumber\\
    +&c_1{\beta_1}^\ast\langle|l_3(t)|l_1(t)\rangle+c_2{\beta_1}^\ast\langle|l_3(t)|l_2(t)\rangle+c_3{\beta_1}^\ast\nonumber\\
	\label{equation 10},
\end{align}
if we only want to get $c_1$ so that we can obtain these two equations

\begin{align}
	{\alpha_1}^\ast&=-\langle|l_1(t)|l_2(t)\rangle-\langle|l_3(t)|l_2(t)\rangle{\beta_1}^\ast, 
	\\
	{\beta_1}^\ast&=-\langle|l_1(t)|l_3(t)\rangle-\langle|l_2(t)|l_3(t)\rangle{\alpha_1}^\ast
	\label{equation 11}.
\end{align}

By solving these two equation, the coefficients can be obtained that

\begin{align}
	{\alpha_1}^\ast&=\frac{\langle|l_1(t)|l_3(t)\rangle\langle|l_3(t)|l_2(t)\rangle-\langle|l_1(t)|l_2(t)\rangle}{1-|\langle|l_2(t)|l_3(t)\rangle|^2},
	\\
	{\beta_1}^\ast&=\frac{\langle|l_1(t)|l_2(t)\rangle\langle|l_2(t)|l_3(t)\rangle-\langle|l_1(t)|l_3(t)\rangle}{1-|\langle|l_2(t)|l_3(t)\rangle|^2}
	\label{equation 12}.
\end{align}

According to the definition of $|l_2(t)\rangle$ and $|l_3(t)\rangle$, we can obtained the coefficients

\begin{align}
	{\alpha_2}^\ast&=\frac{\langle|l_2(t)|l_1(t)\rangle\langle|l_1(t)|l_3(t)\rangle-\langle|l_2(t)|l_3(t)\rangle}{1-|\langle|l_1(t)|l_3(t)\rangle|^2},
	\\
	{\beta_2}^\ast&=\frac{\langle|l_2(t)|l_3(t)\rangle\langle|l_3(t)|l_1(t)\rangle-\langle|l_2(t)|l_1(t)\rangle}{1-|\langle|l_1(t)|l_3(t)\rangle|^2},\\
	{\alpha_3}^\ast&=\frac{\langle|l_3(t)|l_2(t)\rangle\langle|l_2(t)|l_1(t)\rangle-\langle|l_3(t)|l_1(t)\rangle}{1-|\langle|l_1(t)|l_2(t)\rangle|^2},\\
	\\
	{\beta_3}^\ast&=\frac{\langle|l_3(t)|l_1(t)\rangle\langle|l_1(t)|l_2(t)\rangle-\langle|l_3(t)|l_2(t)\rangle}{1-|\langle|l_1(t)|l_2(t)\rangle|^2},
	\label{equation 13}
\end{align}
in the same way.

\section{Instantaneous amplitudes of eigenstates during the loop evolution that encloses two second-order EPs \label{AppendixC}}

Fig.\ref{appendix2} shows the instantaneous amplitudes of eigenstates during the loop evolution that encloses two second-order EPs with different initial states and starting points.

\section{Instantaneous amplitudes of eigenstates during the loop evolution that encloses a third-order EP \label{AppendixD}}

\begin{figure*}
	\centering
	\includegraphics[width=\linewidth]{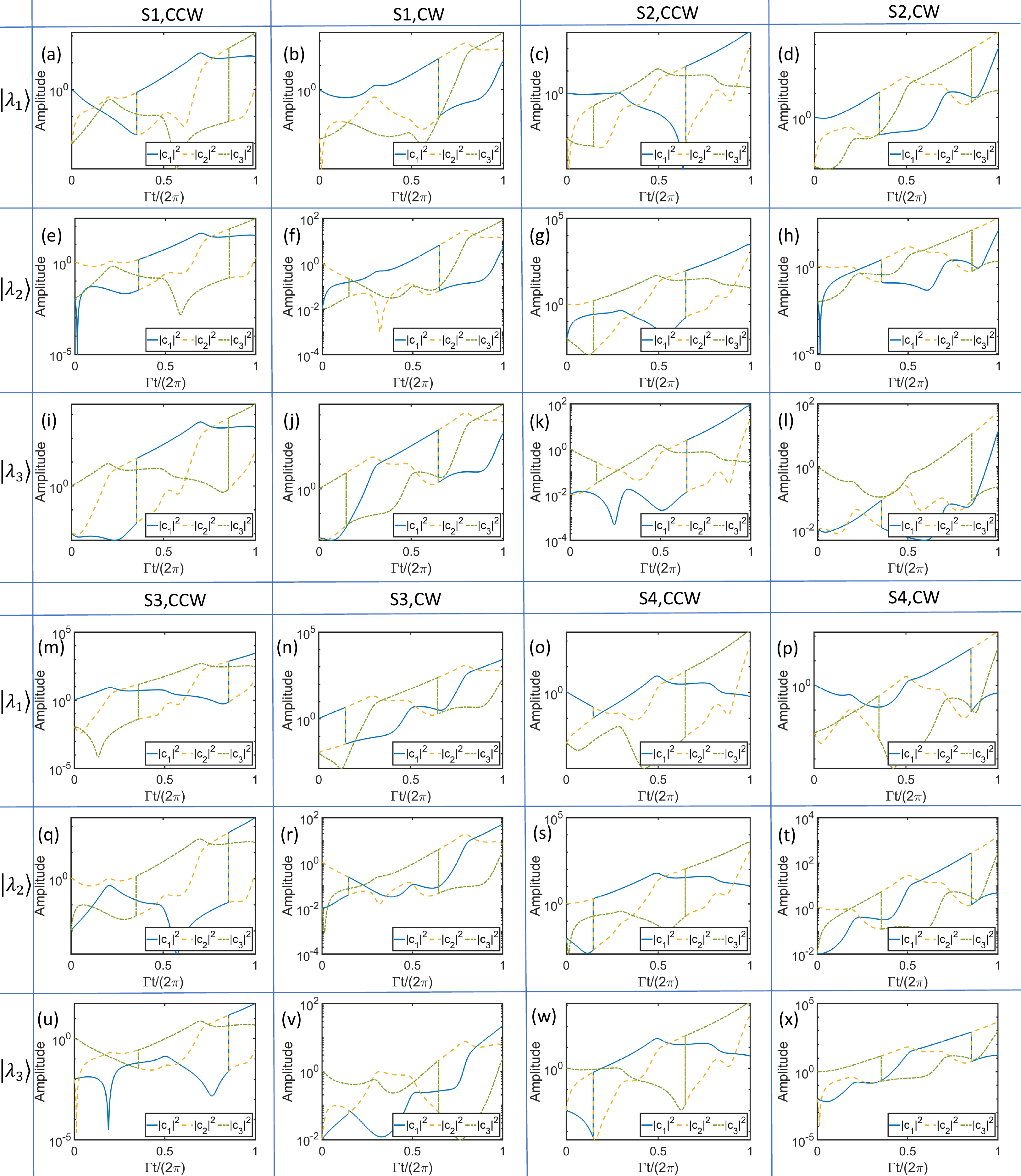}
	\caption{Instantaneous amplitudes of eigenstates during the loop evolution that encloses a third-order EP, where the radius of the loop is 1G. The blue solid line describes the amplitude of $\lambda_1$, the yellow dashed line describes the amplitude of $\lambda_2$ and green dashed dot line describes the amplitude of $\lambda_3$. }
	\label{appendix3}
\end{figure*}

Fig.\ref{appendix3} shows the instantaneous amplitudes of eigenstates during the loop evolution that encloses a third-order EP with different initial states and starting points.

\end{document}